\documentclass[journal=mamobx,manuscript=article]{achemso}
\usepackage[T1]{fontenc}
\usepackage[latin9]{inputenc}
\usepackage{textcomp}
\usepackage{amsthm}
\usepackage{amsmath}
\usepackage{graphicx}
\usepackage{amstext}
\usepackage{bm}

\newcommand{\bq}{\begin{equation}}
\newcommand{\eq}{\end{equation}}

\author{Danijel~Grgi\v{c}in}
\affiliation
{Institut za fiziku, Bijeni\v{c}ka 46, 10000 Zagreb, Croatia}
\email{dgrgicin@ifs.hr}
\phone{+385 98 9523 914}

\title{Probing and evaluating the electric potential of a polyelectrolyte with dielectric spectroscopy}

\abbreviations{DNA, SAXS}

\begin{document}
	\begin{figure*}
		\includegraphics[width=83mm]{TOC.pdf}
		\caption*{Table of Contents}
		\nonumber
		\label{Table of Contents}
	\end{figure*}
	
	\begin{abstract}
		The electromagnetic potential is the only force relevant to understand polyelectrolytes and it should enables us to reveal all polyelectrolytes properties. We argue that dielectric spectroscopy probes the average dipole moment of the pure water polyelectrolyte probing this way the polyion - counterion potential. From this polyion potential can be evaluated in the region from the polyion vicinity all the way to the bulk. 
	\end{abstract}
	\section{Introduction}
	Competition of the electrostatic interaction and entropy of mobile charges causes accumulation of charged counterions to oppositely charged surface of the polyion\cite{naji}. This effect can be comprehended within the framework of the Manning-Oosawa (MO) counterion condensation theory in which counterions accumulate in a vicinity of cylindrical surface of the polyion if the charge density parameter $\eta>1$ ($\eta=l_B/b$, $b=l/N$)\cite{oosawaknjiga71,Schmitz93}. $b$ is the linear charge density, $l$ is the polyion contour length and $N$ is the number of charged monomers. $l_B$ is the Bjerrum length which is the length at which electrostatic interaction between charges equals their thermal motion
	
	\begin{equation}
	l_B=\frac{e_0^2}{4\pi\varepsilon_0\varepsilon_r}\frac{z_1z_2}{kT}.
	\label{Bjerrum}
	\end{equation}
	
	$e_0$ is the elementary charge, $z_1$ and $z_2$ are valences of charged species, $\varepsilon_0$ is the permeability of vacuum, $\varepsilon_r$ is the dielectric constant of the solvent, and $kT$ is the thermal energy. So looking back $\eta$ can also be interpreted as the measure of electrostatic interaction strength relative to the thermal motion. Counterions will accumulate in a condensation layer until they effectively reduce the value of $\eta$ to one, i.e. until they effectively increase separation between the polyion charges from $b$ to $l_B$. The fraction of condensed counterions is given by $1-1/\eta$ where fraction of non condensed counterions is 
	
	\begin{equation}
	f=1/\eta
	\label{free}
	\end{equation}
	
	For small enough charge densities screening of mobile ions in the solution can be described by the Debye-H{\"u}ckel (DH) equation (linearized Poisson-Boltzmann (PB) equation). DH screening is quantified by screening length $\kappa^{-1}$
	
	\begin{equation}
	\kappa^{-1}[nm]\approx 9.63*I_S^{-1/2}[mM].
	\end{equation}   
	
	$I_S=1/2\sum_ic_i[mM]z_i^2$ where $c_i[mM]$ is the concentration of a charged species of valency $z_i$. The meaning of a $\kappa^{-1}$ is very important when visualizing electrolytes since ion would sense solution distanced from it for more than the $\kappa^{-1}$ as electroneutral due to screening from other ions. But the DH approximation is only valid for a system with low charge densities as compared to the kT, i.e. for weakly charged polyions whose linear charge density is much lower than 1 charge per Bjerrum length $(\eta \ll 1)$.
	
	Structure of a semidilute highly charged polyion will be determined by the electrostatic repulsion of charged groups and will locally adapt rodlike structure with cylindrical symmetry. This enables construction of equipotential cylindrical volumes. Importance of constructing equipotential volumes is a possibility to cover all the polyelectrolyte with such volumes. This is in principle enough to model properties of the entire system. Lifson and Katchalsky (L-K) took this approach in 1950's \cite{katchalsky12, katchalsky13} and solve PB equation exactly. They obtain monotonically increasing potential with increasing distance from polyion. That directly produces a monotonic decrease of the counterion radial distribution function. Such approach is not viable when considering added salt solutions. In order to model potential of solution with high added salt ($c>10$mM) approach based on MO counterion condensation theory is used\cite{ray98,ray99}. The potential there is much more complicated as compared to one in pure water solutions.

	For dilute solutions in which distance between the polyions $(\xi)$ is far greater than their contour length $(l)$ equipotential surface is sphere like\cite{dobrynin}. For such equipotential surface the PB equation cannot be exactly solved. To deal with spherical equipotential surface theories which exploit numerical calculation arrived in 2000's\cite{deshkovski, antypov, hansen}. 
	
	Experimental study of colligative properties of polyelectrolytes started with measuring the osmotic pressure \cite{kern,kern2,nagasawa,alexandrowicz,alexandrowicz2}. Osmotic pressure can be linked with electrostatic potential since the prevailing view is that only "free" counterions contribute to osmotic pressure. Experiments showed apparent deficit in osmotic pressure of highly charged polyions $(\eta>1)$ which lead to the conclusion that only "free" counterions contribute to the osmotic pressure and that others are condensed on the polyion. Their mutual ratio was in very good numerical agreement with the MO condensation theory\cite{manning69,oosawaknjiga71,liao}.
	
	Ito et al. showed that there exists a more direct method of probing the p-c potential\cite{ito}. The characteristic length $L$ obtained by dielectric spectroscopy depends on the polyion concentration ($c$); $L\propto c^{-0.33}$ and $L\propto c^{-0.5}$ for dilute and semidilute solution respectively. These dependences are equal to dependencies of the distance between polyions $(\xi)$\cite{deGennes76} so Ito et. al concluded that $L$ measures $\xi$. But the value of $L$ for semidilute solutions is much closer to the value $\kappa^{-1}$ of non condensed counterions than to the value of $\xi$\cite{grgicin2013,salamon}. On the other hand in dilute solution $L\propto c^{-0.33}$ which is the same dependence as one would expect for $\xi$ but not for $\kappa^{-1}$ (which should remain the same as for semi-dilute solution $\kappa^{-1}\propto c^{-0.5}$)\cite{ito,epl08}. So there is still an open question about the interpretation of the $L$, does it measure $\xi$ or $\kappa^{-1}$ or, most probably, neither of them. 
	
	From the 1975. it is known that ds DNA can be denatured to ss DNA by high temperature\cite{record}. We have done a study in which we change the distance between polyions by heat denaturing double stranded (ds) DNA into single stranded (ss) DNA. The measured change of $L$ could be explained by shrinkage of the distance between polyions. Using de Gennes theory we can calculate prediction of the shrinkage

	\begin{equation}
	\frac{\xi_{DS}}{\xi_{SS}}\sim\left(\frac{b_{DS}}{b_{SS}}\frac{c_{DS}}{c_{SS}}\right)^{-0.5}=\left(\frac{3.4}{4.3}\frac{1}{2}\right)^{-0.5}=1.6
	\label{1.6}
	\end{equation}
	
	The change is in very good agreement with one experimentally obtained for the Na-DNA and this also mislead some authors \cite{salamon} to conclude that $L$ is equal to $\xi$.  Unfortunately upon denaturation $\kappa^{-1}$ also changes by the exactly same value as $\xi$.  
	
	\begin{equation}
	\frac{\kappa^{-1}_{DS}}{\kappa^{-1}_{SS}}=\left(\frac{I_{SS}}{I_{DS}}\right)^{0.5}=\left(\frac{f_{SS}\ c_{SS}\ z^2_{SS}}{f_{DS}\ c_{DS}\ z^2_{DS}}\right)^{0.5}=\left(\frac{f_{SS}}{f_{DS}}\right)^{0.5}=\left(\frac{0.62}{0.24}\right)^{0.5}=1.6
	\label{1.6}
	\end{equation}

	With the denaturation experiment one cannot distinguish if $L$ measures $\xi$ or $\kappa^{-1}$ but with the following experiment one can.
	
\begin{figure}
		\includegraphics[width=82.5mm]{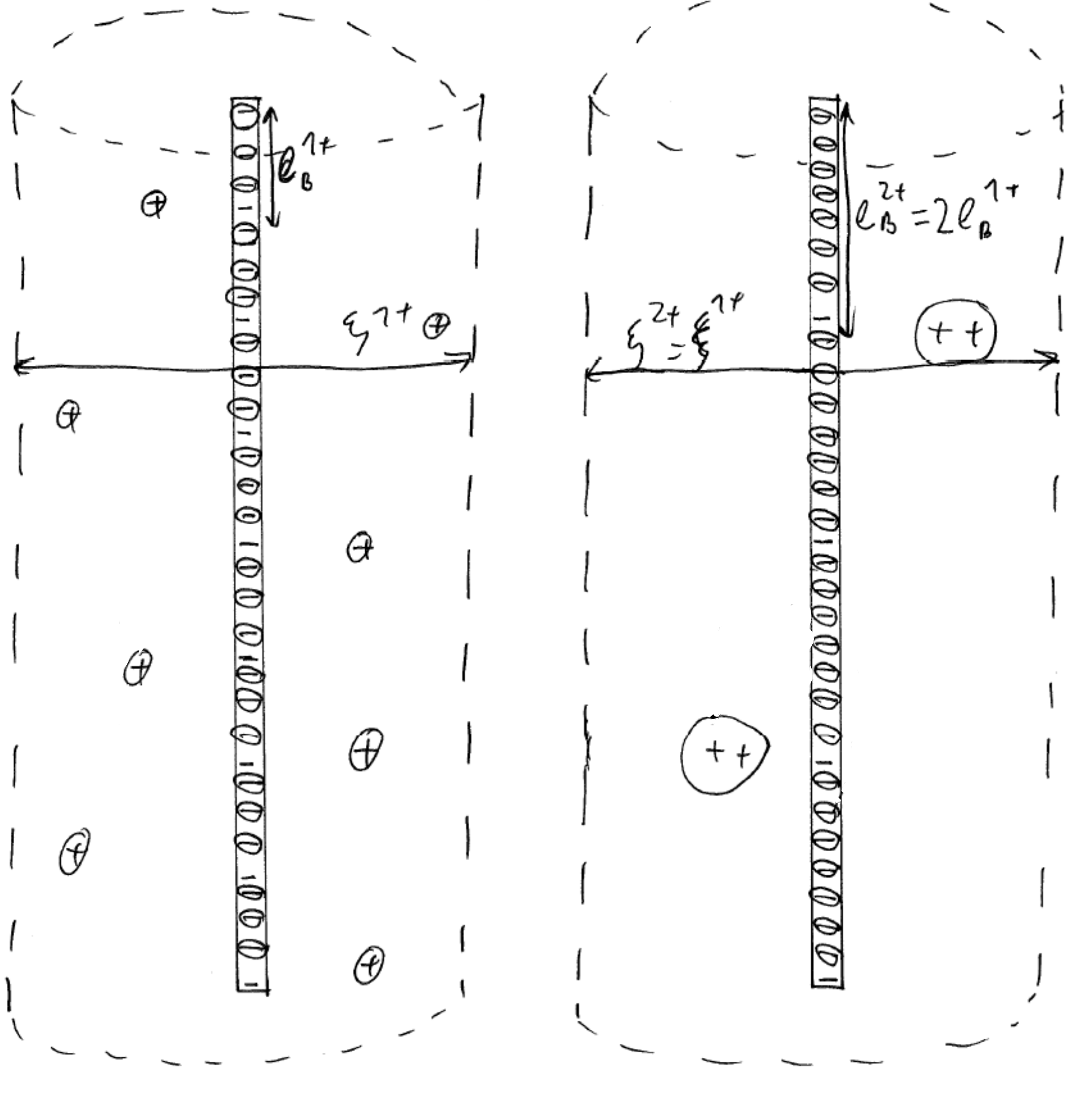}
		\caption{Schematic of pure water polyelectrolyte for polyions with equal concentration but with different counterion valency. Circled polyion charges symbolize neutralization by condensing counterions. $\xi$ is not a function of counterion valency. It has the same value for both polyelectrolytes while $l_B$ and $f$ differ. Number concentration of non condensed divalent counterions is 4 times smaller than that of monovalent.}
		\label{fig1}
	\end{figure}

	For two monovalent ions in solution at 25$^o$C $l_B=0.72\ nm$ but for one monovalent and one divalent ion $l_B$ doubles to $l_B=1.44\ nm$. That has a consequence on condensation. Two times larger $l_B$ causes two time smaller effective charge per polyion length, i.e. $1e^-/1.44\ nm$ instead of $1e^-/0.72\ nm$ since counterions condense on the polyion until they reduce the charge to one per $l_B$\cite{combet}. Since more of counterions are condensed less of them are non condensed (situated in the bulk further away from polyion). In the pure water sodium DNA solution 24\% of the sodium counterions are non condensed $f(Na^+)=0.24$ while in the solution with magnesium counterions only 12\% of the counterions are non condensed $f(Mg^{2+})=0.12$. Although the number of non condensed monovalent charges would be only 2 times larger the number concentration of non condensed monovalent counterions would be 4 times larger as compared to divalent ones for the same concentration of polyions, see Fig. \ref{fig1} and table \ref{tab1} for clarification.
	
	\begin{table}
		\centering
		\caption{Linear charge density $b$, Bjerrum length for corresponding counterions $l_b$ and fraction of non condensed counterions $f$ for DNA, hyaluronic acid (HA) and poly(styrene sulphonate) (PSS).}
		\begin{tabular}{ccccc}
			\hline 
			&$b(e^-/nm)$&$l_b (nm)$ & $f$  \\
			\hline
			Na-DNA & 0.17 & 0.72 & 0.24 \\
			ss Na-DNA & 0.43 & 0.72 & 0.60\\
			Mg-DNA & 0.17 & 1.44 & 0.12 \\		
			Na-HA & 1 & 0.72& 1 \\
			Mg-HA & 1 & 1.44& 0.70 \\
			Na-PSS & 0.25 & 0.72 & 0.35\\
			\hline				
		\end{tabular}
		\label{tab1}%
	\end{table}
	Polyelectrolytes used in this study were chosen to have different values of $l_b$ ($f$). $\xi$ we have regulated with polyion concentration. Obtained data suggest that Ito's interpretation, whose many studies had used\cite{pre07,grgicin2013,norio,saif,litovitz,lee,gabriel,hayakawa,bakewell,omori,katsumoto,elena} cannot be a valid interpretation for measured $L$.

	\section{MATERIALS AND METHODS}
	
	Gel electrophoretic analysis of lyophilized salmon testes Na-DNA threads (Sigma Cat. No. D1626) revealed polydispersity in the range of 1-20 kbp \cite{Grgicin2018}. This correspond to the average polyion length of $3.5\ \mu m$ since one base pair height is $3.4$ \AA. Here we report results of DS on a DNA, Hyaluronic Acid (Sigma Cat. No. 53747) (HA) and poly(styrene sulfonate) (Polymer StandardS Service GMBH, psskit-07, Mp=679000, Mw=666000) aqueous solution with monovalent Na$^+$ and divalent Mg$^{2+}$ and Mn$^{2+}$ counterions. Pure water Na-DNA/Na-HA/Na-PSS solution was produced by dissolving Na-DNA threads/Na-HA powder/Na-PSS powder according to the protocol I. from Ref.~\citenum{pre07}. DNA/HA with Mg$^{2+}$ or Mn$^{2+}$ counterions were produced by dissolving Na-DNA threads/Na-HA powder in MgCl$_2$ or MnCl$_2$ salt of higher concentration than concentration of the $Na^+$. Then we dialyzed obtained solutions against the MgCl$_2$ or MnCl$_2$ in order to completely get rid of a $Na^+$ counterions. After that we dialyzed solution against pure water in order to get rid of excess MgCl$_2$ or MnCl$_2$, i.e. to obtain solutions without added salt. Finally, samples in the concentration range $0.01g/L<c_m<5g/L$ were done according to the protocol I. from Ref.~\cite{pre07}.      
	Except the Na-PSS, the samples with the lowest produced concentration of $0.01g/L$ was above de Gennes dilute-semidilute crossover value which means that all produced samples were in semidilute regime. For them distance between polymers are equal to de Gennes correlation length $\xi\propto (bn)^{-1/2}$. Here $n$ is the number concentration of monomers\cite{deGennes76}. Sometimes it is more appropriate to use molar concentrations of phosphate/monomers  $c[mM]=c_m[g/L]*3\ mmol/g$, $c[mM]=c_m[g/L]*2.5\ mmol/g$ and $c[mM]=c_m[g/L]*4.85\ mmol/g$ respectively for DNA, HA and PSS.
	
	Native dsDNA consists of two complementary strands so it can be denatured to two single strands (ss). Denaturation was done by exposing samples to temperature of 97$^{\circ}$C for 20 minutes. This temperature and duration were enough to cause denaturation of DNA\cite{record}. All measurements are done at 25.00$\pm 0.01^{\circ}$C and temperature was controlled with Peltier elements. ssDNA monomer (phosphate) is longer than a base pair in dsDNA form, $4.3$\ \AA $ $ against $3.4$\ \AA $ $ respectively. Based solely on the de Gennes distance between polyions $\xi\propto (bn)^{-1/2}$ this variation in $b$ and twice larger $n$ gives 1.6 times shorter distance between the polyions upon denaturation. That is very similar to the experimentally obtained value. 
	
\begin{figure}
		\includegraphics[width=82.5mm]{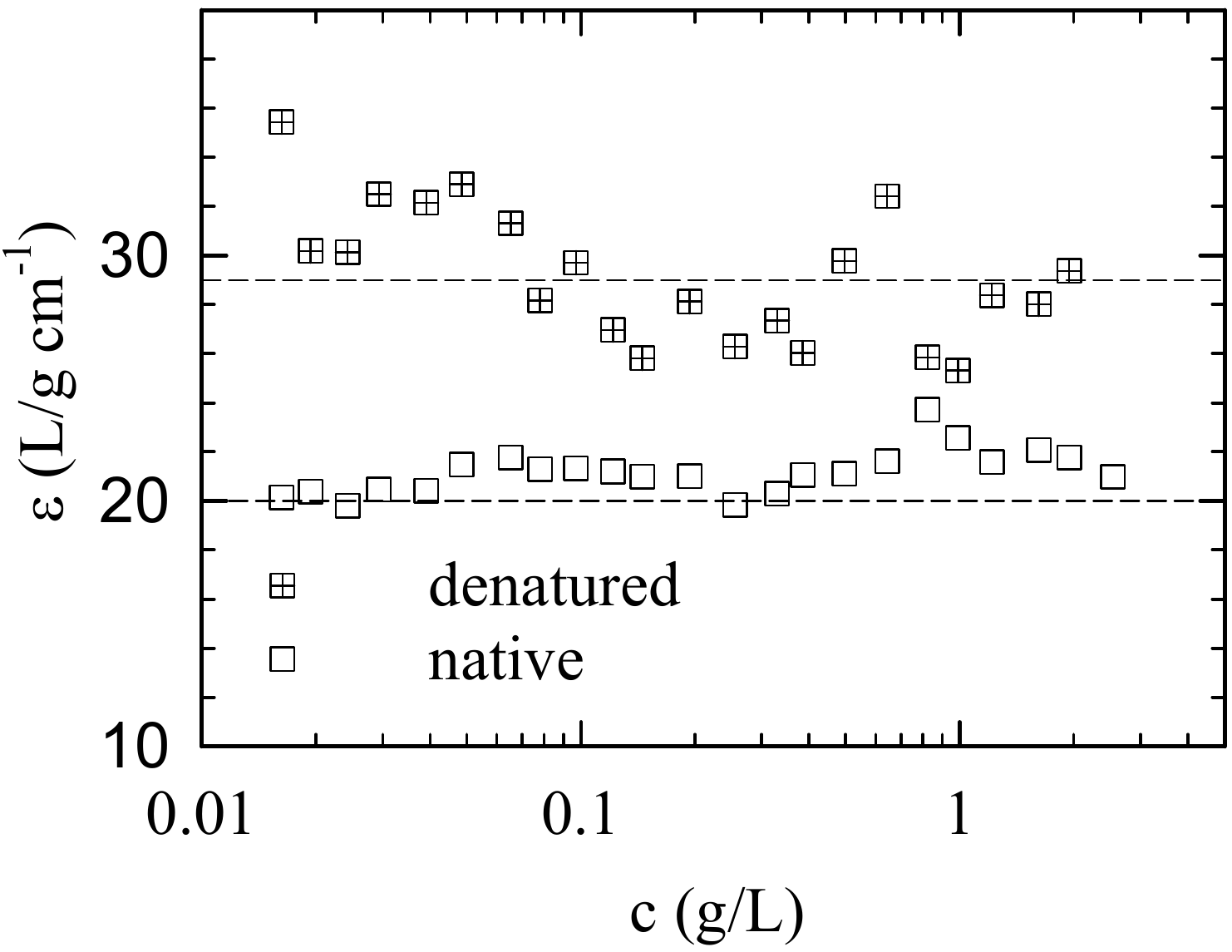}
		\caption{Extinction coefficient of native (squares) and denatured (crossed squares) Mg-DNA as a function of concentration. Dashed lines represent expected ds and ss values for extinction coefficient.}
		\label{figuv}
	\end{figure}

	The problem of renaturation can arise if one cools a sample too slowly. If given enough time a ssDNA in a solution at the low enough temperature will partially renature. Time needed for renaturation is longer if the strands are longer and if the solution temperature is lower. On the higher temperature strands "search" faster through the solution and have more chance to find a homologous pair on another strand. So if one cools denatured samples quicker the time window in which temperature is low enough for renaturation to occur and high enough to facilitate pairing of strands is shorter. Although renaturation occurs also on 25$^{\circ}$C but it is not significant. Nevertheless, we keep the time till measurement is done as low as possible. Typically, measurement of denatured sample is done under 10 minutes after the sample was taken from mixture of ice and water. Signature of renaturation can be seen from UV spectrophotometric data although we have taken all precautions. Data of denatured Mg-DNA, shown on Fig. \ref{figuv}, deviate much more from expected value than data of native samples.
	
	UV spectrophotometry enables determination of the conformation of DNA in a solution since ssDNA absorbs around 40\% more electromagnetic radiation than dsDNA of the same concentration at 260 nm \cite{bloomfield}. For a 0.05 g/L dsDNA we choose absorption to be one (A=1) and that fix extinction coefficient value to $\varepsilon=20 L/gcm$ for a dsDNA and around $\varepsilon=29 L/gcm$ for a ssDNA. More about UV spectrophotometry can be found in our previous publication\cite{hazu}.
	
	Our DS technique uses in house built capacitive chamber whose distance between parallel plates is $l=0.1021\pm 0.0001cm$ and the chamber constant corresponding to the sample volume of $100\ \mu L$ is the $l/S=0.1042\pm 0.008cm^{-1}$. The capacitive chamber enables reliable complex admittance measurements with reproducibility of 1.5\% of sample with conductivities in the range $1.5-2000 \mu S/cm$\cite{pre07}. With the same chamber we can also obtain another electrode separation $l=0.1558\pm 0.0001cm$. With the Agilent 4294A impedance analyzer we sweep through frequency range $\nu=100\ Hz-15\ MHz$. Ac amplitude of the $50\ mV$ was chosen to probe samples after establishing that measurements are independent of applied voltage in the range from 5-1000 mV. Measurement of all samples ($0.01g/L<c<5g/L$) including preparation of $5 g/L$ mother solution was done within 12 hours. More detailed description of technique can be found in our previous publications\cite{pre07,vuletic10,physicab}. 
	
\begin{figure}
		\includegraphics[width=82.5mm]{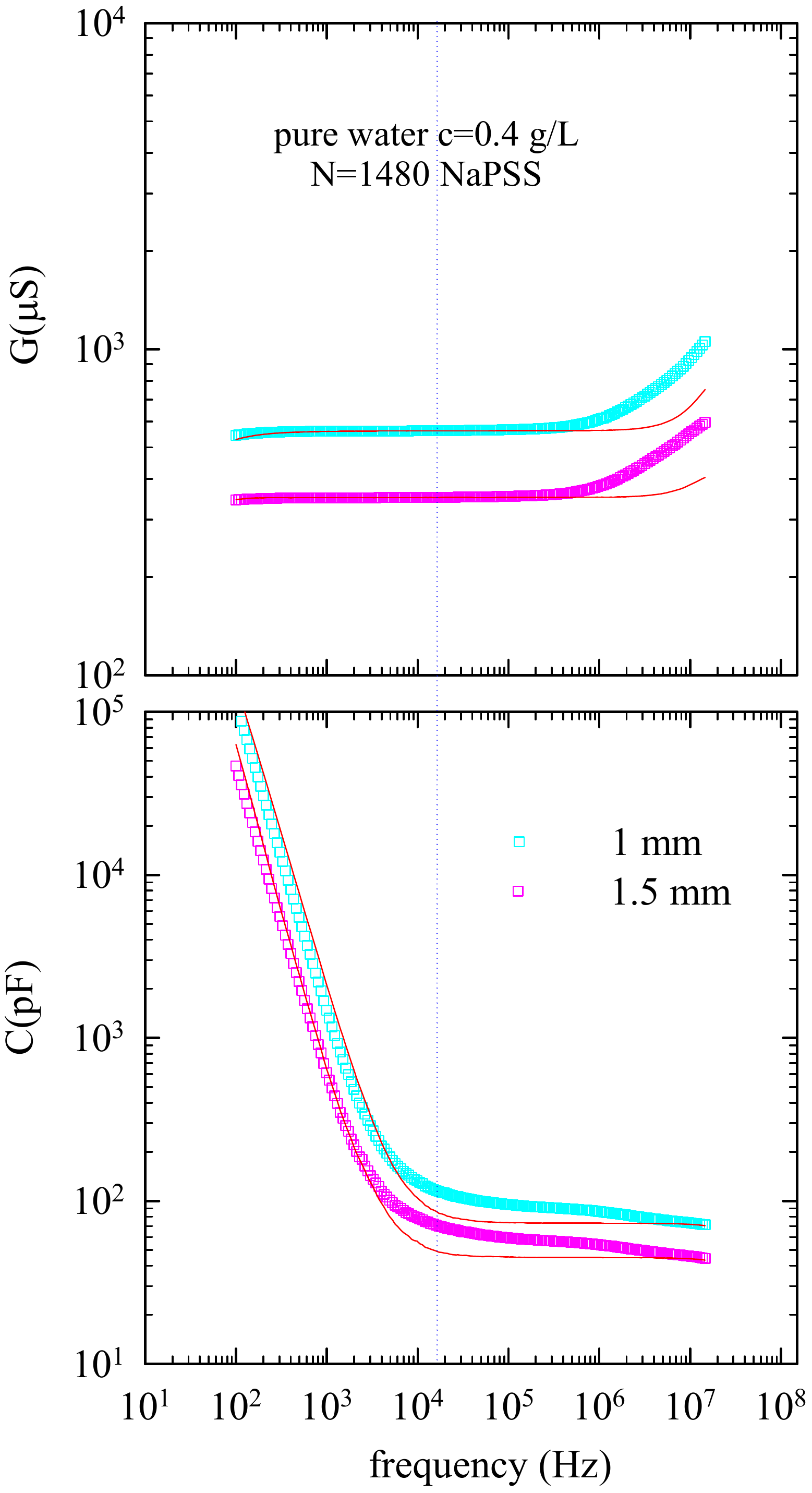}
		\caption{$G$ and $C$ spectra of pure water NaPSS solutions (N=1480) of 0.4 g/L concentration (open symbols) and electrolyte (red line). Blue dotted lines represent frequency up to which electrode polarization effects dominate measured data.}
		\label{rawspektri}
	\end{figure}

	Since we model our equivalent circuit as parallel our raw data consist of the conductance $G(\omega)$ and capacitance $C_p(\omega)$. Increment of dielectric function we obtain as 
	\begin{equation}
	\Delta \varepsilon = \frac{l}{S}\left(\left(\frac{C_{sample}(\omega)-C_{ref}(\omega)}{\varepsilon_0}\right)-i\left(\frac{ G_{sample}(\omega)-G_{ref}(\omega)}{\varepsilon_0 \omega}\right)\right)
	\label{increment}
	\end{equation}
	
	where $\varepsilon^*$ can be obtained as $\varepsilon^*=\Delta\varepsilon +\varepsilon_{\infty}$, where $\varepsilon_{\infty}$ is the high-frequency limit of $\varepsilon^*$. On Fig. \ref{rawspektri} we show raw $G$ and $C$ spectra for typical polyelectrolytes and reference electrolyte.
	
\begin{figure}
		\includegraphics[width=82.5mm]{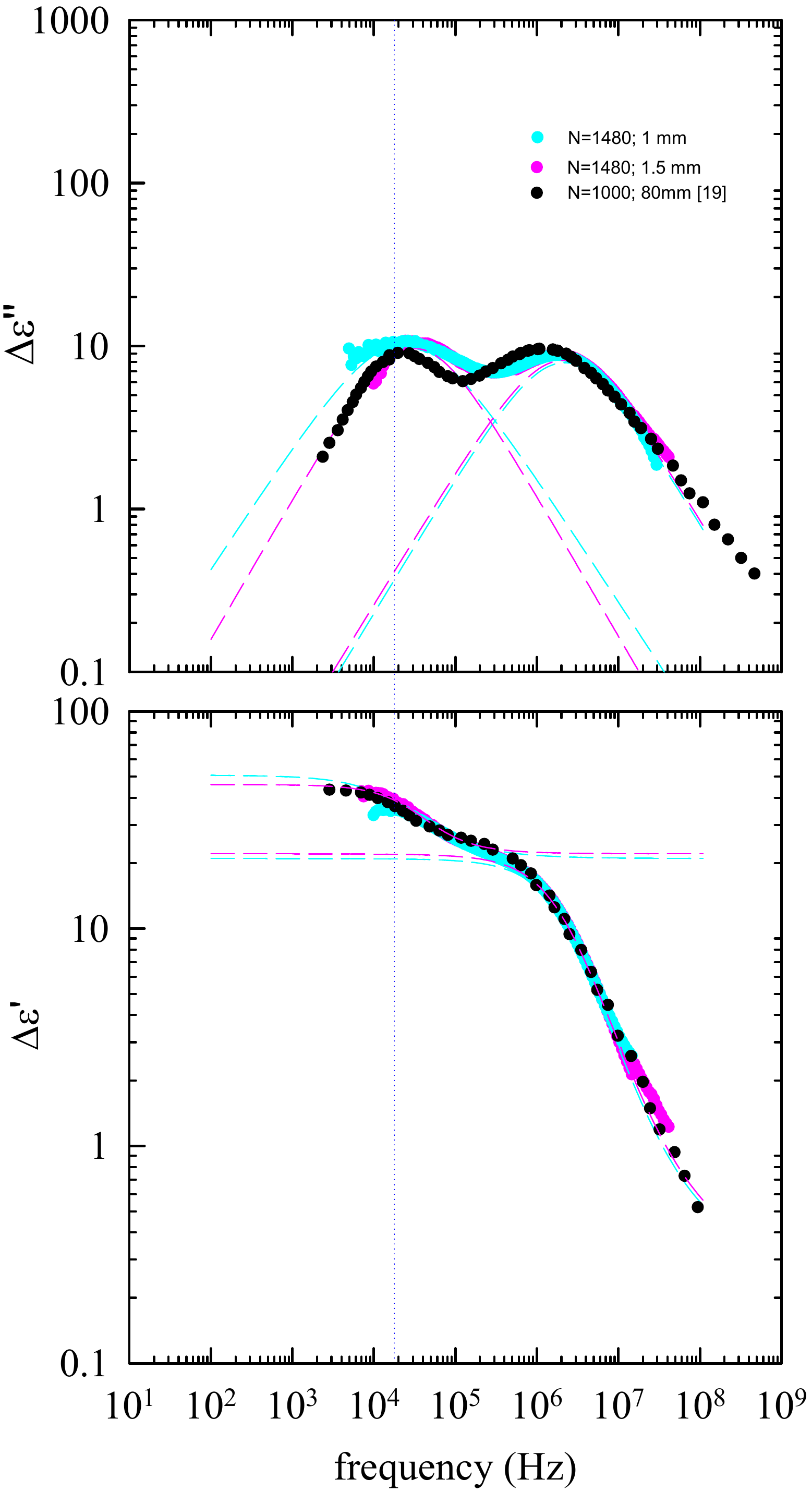}
		\caption{Imaginary $(\Delta \varepsilon'')$ and real $(\Delta \varepsilon')$ increment of dielectric function of the same polyelectrolytes as on Fig. \ref{rawspektri} with addition of N=1000 data from reference \citenum{ito}. Full lines are fits to the sum of two Cole-Cole functions while dashed lines represent single Cole-Cole fit. Dotted blue lines is on the same frequency as on Fig. \ref{rawspektri}.}
		\label{spektri}
	\end{figure}

	Since reference electrolyte does not have relaxation for which polyions are needed it is easier from their spectra to determine frequency up to which electrode polarization dominates experimental data. At the frequency of $2*10^4\ Hz$ and above capacitance $C$ starts to be frequency independent and that tells us that electrode polarization does not dominate experimental data at those frequency. On the Fig. \ref{rawspektri} we indicate that frequency with vertical blue line. There exist other procedures\cite{Bordi} of dealing with electrode polarization, each with upsides and downsides: 4-contact measurement cells \cite{Schwan68}, black platinum electrodes\cite{Holling}, modeling the electrode polarization contribution\cite{Roldan, Klein}, measurements at different electrode separations \cite{Holling}. Electrode polarization depends on many variables non linearly so modeling or reference subtraction method \cite{Davey,Grosse} cannot perfectly address electrode polarization. Because of that we use combination of reference subtraction method and measurement at different electrodes separations (0.1 cm and 0.155 cm). 	
	$C$ of NaPSS samples are only a little higher than value of corresponding added salt. So if there is some relaxation caused by polyion in this frequency range it is of small dielectric strength. In order to detect it as a reference solution we choose electrolyte with similar conductivity as of polyelectrolyte sample of interest. In that way we would be able to extract relaxation caused by polyion/polyelectrolyte since our reference sample should not contain it. By such simple subtraction we obtain increment of dielectric function $\Delta \varepsilon $ of polyelectrolyte as compared to electrolyte. Such measurement subtraction method are followed by comparison of increments of dielectric functions obtained with two different electrode separations which we show on Fig. \ref{spektri}. We see data starts not to differ at exactly  $2*10^4\ Hz$ which we previously recognize as frequency up to which electrode polarization dominates data. Next we see that parameters of lower frequency relaxation has large uncertainties and that parameters of higher frequency relaxation are almost identical. Because of that we decided not to further process lower frequency data and process only higher frequency data for which we prove that they are not dominated by electrode polarization. Further evidence towards that conclusion provides results of Ito et. al obtained on electrode separation of 80 mm\cite{ito komora} which higher frequency relaxation slightly difference from ours can be ascribed to sample difference (see SI).
	Next, to further address error bars of our measurement we made 3 independent experiments on NaPSS $N=3300$ sample all the way from the sample preparation to data analysis and our data are within error bars identical to Ito's et al data for NaPSS $N=3800$. 
	Demonstrating precision with this three measurements shown on Fig. \ref{fig4} with different colors we prove that $L$ of polyions with divalent counterion significantly differs from the $L$ of polyions with monovalent counterions.
	
	\section{RESULTS}
	Within phosphate concentration range $0.08\ mM<c<1\ mM$ typical DS responses of pure water Mn-DNA solutions with divalent counterions are shown in Fig.\ \ref{spektri}. Counterions collectively respond through two processes, one centered between $10^3-10^5\ Hz$ and the other between $10^5-10^7\ Hz$. In order to analyze these processes we fit a sum of two Cole-Cole functions which both have 3 parameters; mean relaxation time $\tau_0$, dielectric strength $\Delta \varepsilon$ and symmetrical broadening of relaxation time $1-\alpha$. Characteristic length of relaxation $L$ can be obtained from mean relaxation time $\tau_0$ as $L=\sqrt{D_{in}\tau_0}$ where $D_{in}$ is the diffusion coefficient $(1.334,\ 0.706,\ 0.712)*10^{-9}m^2/s$ respectively for Na, Mg and Mn counterions\cite{Lide}). Assumption that polyion diffusion is negligible compared to counterion is justified since DNA polyion has much smaller diffusion constant\cite{salamon}. In this report we will focus only to $10^5-10^7\ Hz$ relaxation. 
	
\begin{figure}
		\includegraphics[width=82.5mm]{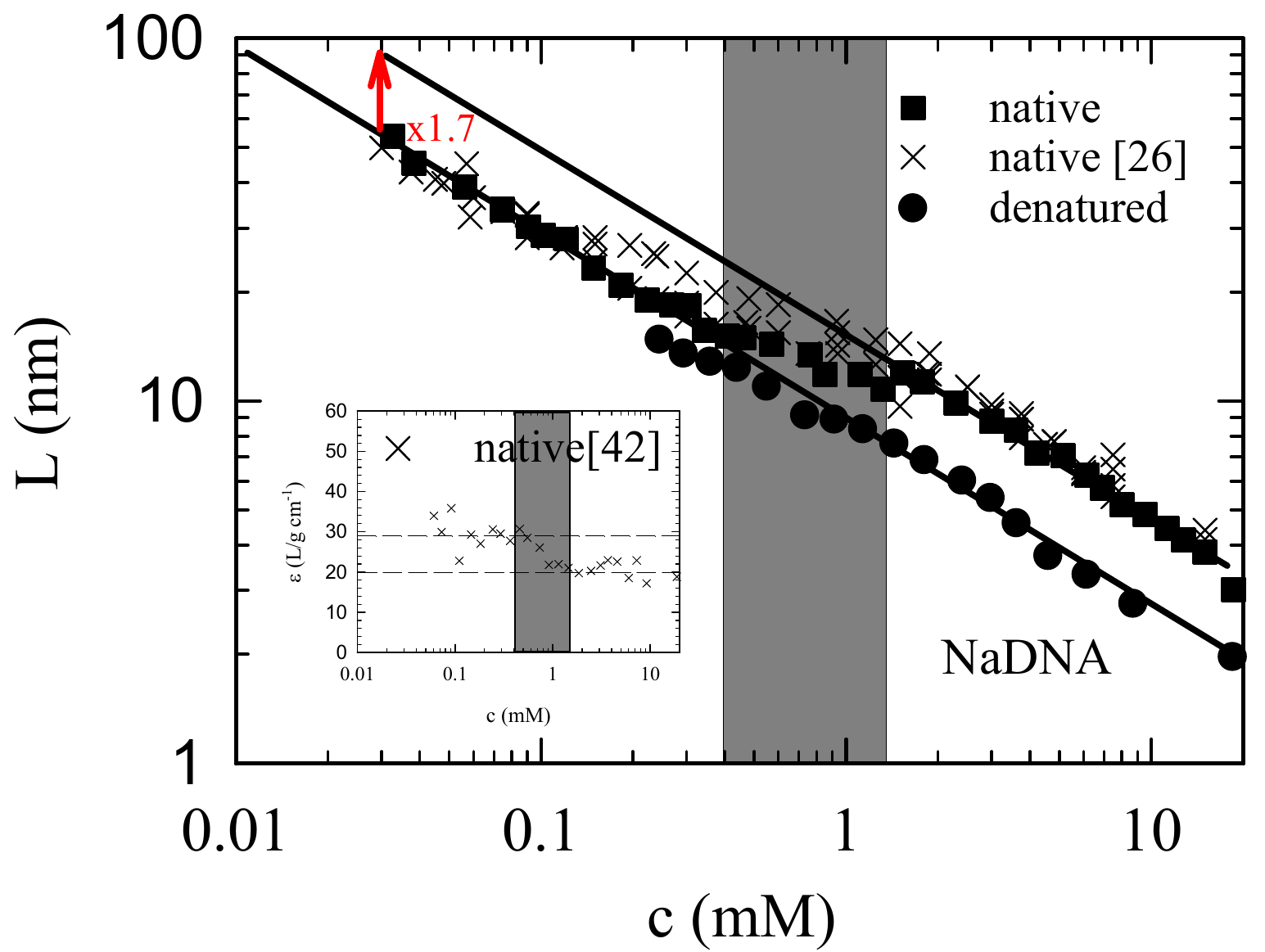}
		\caption{Characteristic length $L$ of pure water Na-DNA as a function of concentration of phosphates for native/denatured samples. Both black lines show $L\propto c^{-0.5}$ dependence. Value in top left corner indicate by how much black lines differ. Grey area corresponds to transition concentrations in which native Na-DNA transit from one to another $L\propto c^{-0.5}$ line. Inset: Extinction coefficient of native Na-DNA as a function of concentration, data taken from Ref. \citenum{hazu}.}
		\label{fig3}
	\end{figure}

	Although DS response of Na-DNA in pure water solution was already published\cite{pre07} open questions remained regarding nature of counterion relaxation, i.e. underlying p-c potential. Is it governed by concentration of polyions ($\xi$) or polyion linear charge density $(b)$ ? In Fig. \ref{fig3} we show $L$ versus concentration of phosphates for native and denatured Na-DNA together with previously published native Na-DNA data.
	
	It can be seen that all measured data for both native and denatured DNA's lie on one of $L\propto c^{-0.5}$ lines except data of native Na-DNA samples in the concentration range $0.04\ mM<c<1.5\ mM$. In this range one can observe the transit from one to the other line which corresponds to a transit from ds to ss conformation of DNA. This transit was additionally confirmed by UV spectrophotometry.	Data of denatured samples deviate slightly more from corresponding line then data of native samples. That is due to the much more complicated preparation procedure and renaturation of strands. After denaturing DNA at 97$^{\circ}$C one must cool down the sample to measuring temperature (which in our case was 25$^{\circ}$C) as quickly as possible. Even that is not the guarantee for avoiding renaturation. $L$ of native samples are equal or slightly above the value for denatured samples on concentrations lower than transitional. That might suggest uncomplete denaturation of native samples on lower concentration. As dsDNA denatures into two strands the $\xi$ should shrink.   
	
	\begin{figure}
		\includegraphics[width=82.5mm]{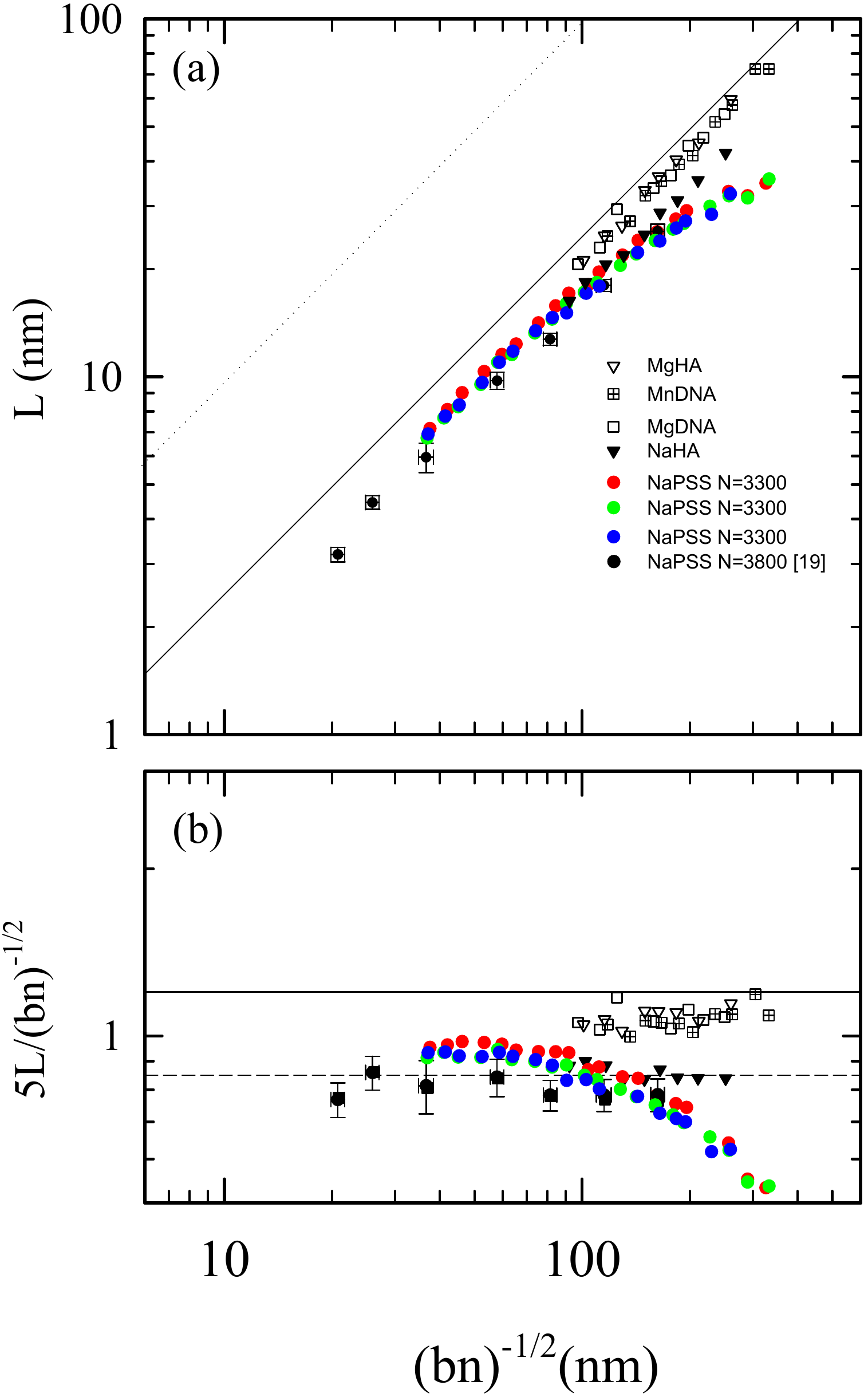}
		\caption{Characteristic length (a) $L$ and (b) normalized $L$ as a function of $(bn)^{-1/2}$ for different polyions (DNA, HA, PSS) and with different counterions (Na, Mg, Mn) in pure water solution. Dotted line represent $\xi=(bn)^{-1/2}$, full lines are our model prediction of $L$ for semidilute solution and dashed line is guide for an eye.}
		\label{fig4}
	\end{figure}
	
	\noindent
	On Fig. \ref{fig4} we show $L$ vs $\xi$ for different polyions with different counterions. Firstly we see that $L$ is shifted from the $\xi$ (diagonal dotted line). This shift is around the factor of 5. That value is similar as in previously published data\cite{ida}. In that paper authors found that $2\pi L$ corresponds to $\xi$. They speculated $2\pi$ is simply due to a factor in Einstein-Smoluchowski equation. All polyelectrolytes on Fig. \ref{fig4} are semidilute except NaPSS for $(bn)^{-1/2}>100$ which are dilute. That is the reason why their value differs from the values of all semidilute polymers and that is known for 30 years. Fig. 3(a) from the reference \citenum{ito} clearly showed that (more information available in SI). 
	
	In is easy to calculate $\xi$ in semidilute solutions since intra-polyion and inter-polyion distance are the same $\xi=(bn)^{-1/2}$. For dilute solution they are different. They have bimodal distribution of values since it has smaller value for distance between chain segments of the same chain and larger value for distance of two chains. So for semidilute solution one can use single value for $\xi$ but for dilute solution one must use distribution of $\xi$ to calculate $L$.	 Thus is the reason why on the Fig. \ref{fig4} we only show prediction of our model for semidilute Na-DNA and Mg-DNA. The drop of $L$ for NaPSS as it enters dilute solution is expected since $\xi$ of dilute solution is smaller as compared to the semidilute solution. This is known from 1990.\cite{ito}.
	Secondly, polyelectrolytes with monovalent counterions have $30\%$ lower $L$ when compared with divalent counterions. That can be clearly seen in the range $90<(bn)^{-1/2}<330$. This must be somehow linked with electrical property of polyelectrolytes. Polyelectrolyte linear charge density $b$ differs among these polyelectrolytes and we argue that it influences p-c potential which in turn influences counterion movement. In order to explain our experimental data we propose the following model

	\noindent	
	
	\begin{itemize}
		\item Polyion is modelled as a hard cylinder with line charge in its center. Linear density of line charge was taken as one predicted with MO condensation theory\cite{deserno}. Potential outside the hard cylinder is of form $V=\frac{\lambda}{2\pi\varepsilon_0\varepsilon_r}ln\left(\frac{r}{\xi/2}\right)$ where $\lambda=1/0.72nm$ and $\lambda=1/1.44nm$ in the case of Na-DNA and Mg-DNA respectively. $r$ is the distance from the center of the cylinder which can take the value in the range from $r_a$ to $\xi/2$ where $r_a$ is radius of polyion which is for DNA taken to be $r_a=1nm$.
		\item (Radial) distribution function $g(r)$ is obtained from potential as $g(r)=exp(-V(r))$. The following is the normalized form of $g(r)$ that is used in calculations
		
		\begin{equation}
		g(x)=\frac{\left(x\right)^{-\lambda/(2\pi\varepsilon_0\varepsilon_r)}}{\int_{r_a/\xi}^{1}\left(x\right)^{-\lambda/(2\pi\varepsilon_0\varepsilon_r)}dx}
		\label{g}
		\end{equation}
		\begin{figure}
			\includegraphics[width=82.5mm]{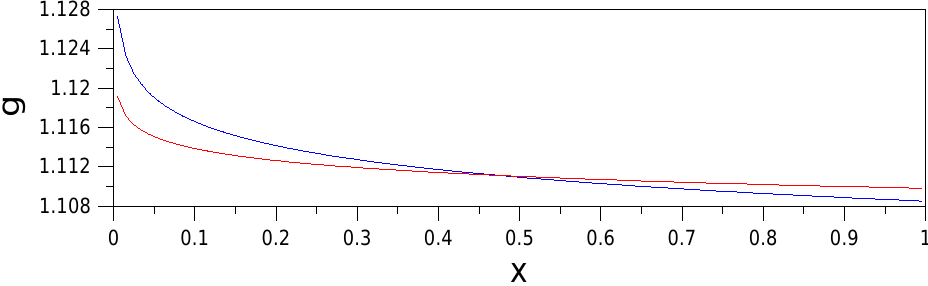}
			\caption{Counterion distribution function calculated with Eqn. \ref{g} for Na-DNA (blue) and Mg-DNA(red) as a function of distance normalized to $\xi/2$.}
			\label{figg}
		\end{figure}

		\item Characteristic length $L$ measured by DS is actually polyelectrolyte average dipole moment $<\mu>$ which is in a static electric field $E$ given by\cite{oosawaknjiga71}
		\begin{equation}
		<\mu>=\frac{\int \mu* exp[-f(\mu)/kT+\mu E/kT]d\mu}{\int exp[-f(\mu)/kT+\mu E/kT]d\mu}
		\label{mu}
		\end{equation}
		$f(\mu)$ is the free energy of the system in a state with a dipole $\mu$. First let us explain the reason why our experimental results do not depend on E. Most of our measured $L$ were smaller than $\mu=50 nm$ and we used the electric field of $E=50V/m$. With these numbers the second factor in the exponent of the Eqn. \ref{mu} has a value of $\mu E/kT=1/1000$. On Fig. \ref{dipol energy} we compare this value with the value of $I=-f(\mu)/kT$. For all values of $x>0.2$ it is smaller. It is almost two orders of magnitude smaller for large values of $x$ which contributes more to the integral in Eqn. \ref{mu}. Thus is the reason our experimental data did not depend on $E$. But more importantly this would allow us to omit this term from our calculation of the average dipole moment.
		
		\begin{figure}
			\includegraphics[width=82.5mm]{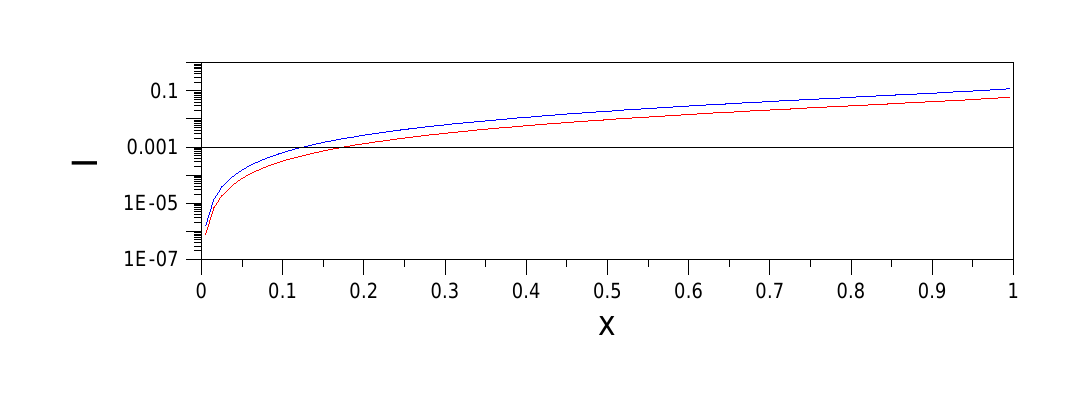}
			\caption{Value of $-f(\mu)/kT$ as a function of distance normalized to $\xi/2$ for Na-DNA (blue) and Mg-DNA (red).}
			\label{dipol energy}
		\end{figure}

		\item
		\begin{figure}
			\includegraphics[width=82.5mm]{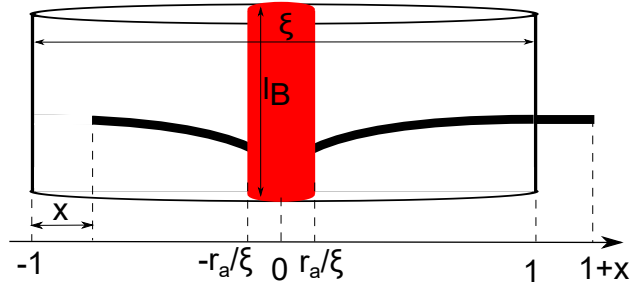}
			\caption{Counterion redistribution facilitated by electric field $E$ expressed as normalized cell radius $0<x<1$. Height of cell (cylinder) is chosen in a way that there is only one charge in the cell (height is equal to $l_B$). That means height is $0.72nm$ and $1.44nm$ for Na-DNA and Mg-DNA respectively.}
			\label{potential}
		\end{figure} 
		
		Free energy of a Na-DNA system in a state with a dipole $\mu$ is calculated as $f(\mu)=\int_1^{1-x}Vq(x)dx$. We need to express the dipole moment as a function of $\mu(x)$. Dipole moment is given by $\mu=\int xg(x) d^3x$. In the evaluation of free energy we made two assumptions. On Fig. \ref{figg} we plotted the value of $g(x)$ obtained using Eqn. \ref{g}. It can be seen that distribution function only slightly depends on position $x$. In the case of Na-DNA its value varies less than 4\%. In the case of Mg-DNA its variation is even lower. That enables us to use position independent counterion distribution function $g=const$. That is in accord with theoretical work of Deshkovski et al.\cite{deshkovski}. For polyelectrolyte they showed that number of localized counterions in 3D space does not depend on r for a saturated condensation phase. The second assumption was made on free energy. We neglect free energy contribution of charge redistribution in the position from 1 to 1+x. Taken these two assumptions we obtain $f(\mu)=2Aq\int_1^{1-x}Vdx=2Aq\int_1^{1-x}lnx\ dx=q2A[(1-x)ln(1-x)+x]$ where $A=1.583*10^{-3}V$.
		
		\begin{equation}
		\mu(x)=\int_{r_a/\xi}^{1-x}xgdx(-\hat{x})+\int_{r_a/\xi}^{1+x}xgdx(\hat{x})=2gx\hat{x},
		\label{mu2}
		\end{equation}  
		were $g=\xi e_0/(4*(1-r_a/\xi))$.
		
		\item Now we combine all those separately obtained pieces into the final expression for average dipole moment of Na-DNA
		\begin{equation}
		<\mu>=\frac{\xi e_0}{4(1-r_a/\xi)}\frac{\int_0^{1-r_a/\xi} 2x* exp(\frac{-q2A[(1-x)ln(1-x)+x]}{kT})dx}{\int_0^{1-r_a/\xi} exp(\frac{-q2A[(1-x)ln(1-x)+x]}{kT})dx}
		\label{final}
		\end{equation}	 
		In order to evaluate integrals from expression \ref{final} we need to specify value of $\xi$. We specify values which correspond to concentration of Na-DNA used in our experiments. For all of them we obtain following result:
		
		\begin{equation}
		L=<\mu>/e_0=0.25\xi
		\label{finalmu}
		\end{equation}
		The same result we obtained for Mg-DNA. This result is valid for semidilute as well as for dilute solutions and with full lines on Fig. \ref{fig4} we showed its prediction for semidilute solution. NaPSS with N=3300 around $(bn)^{-1/2}=100\ nm$ enters dilute regime and that is the reason for their departure from the line for $(bn)^{-1/2}>100\ nm$.

	\end{itemize} 
	
	\section{DISCUSSION}
	
	Since 1990's the dominant interpretation of $L$ for a semidilute solution is that it spans exactly the same distance as $\xi$\cite{ito}. For both dilute and semidilute solutions $L$ has the same concentration dependence as $\xi$ and change in concentration dependence are at similar values as theoretically expected for dilute to semidilute crossover. That leads many to believe that $L$ measures $\xi$\cite{ito,pre07,grgicin2013,norio,saif,litovitz,lee,gabriel,hayakawa,bakewell,omori,katsumoto,elena,salamon}. Actually, $\xi$ can be obtained from characteristic wave-vector $(q_m)$ from small angle X-ray scattering (SAXS) intensity\cite{combet}. It is in perfect agreement with theoretically predicted by $\xi = (bn)^{-1/2}$ and it is 6 times larger than $L$ measured by DS\cite{salamon,ida} suggesting that $L$ is not a measure of $\xi$ 
	
	Using counterions of two different valences we saw a discrepancy which cannot be explained with the existing model because $\xi$ should not be ion dependent and our measured $L$ is ion dependent. We proposed a new model for characteristic length obtained from dielectric spectroscopy of pure water dilute and semi-dilute solutions. We argue that it measures the average dipole moment $<\mu>$. In order to calculate the average dipole moment one needs to know only one variable; polyion-counterion potential of mean force. From it counterion distribution function can be obtained. 
	
	We modeled the polyion with effective line charge situated in the center axis of a hard cylinder with radius which exactly encapsulates the real polyion. This line charge is equal to polyion charge if polyion is weakly charged while in the case of strongly charged polyion effective charge is given by MO condensation theory. With few assumptions that greatly simplify the calculation but all of which we justified either with experimental data or calculus, or both, we get good agreement of calculated/modeled $<\mu>$ with our experimentally obtained $L$. It is interesting to notice that the agreement is perfect in the case of divalent DNA/polyions but slightly less good in the case of monovalent DNA. The reason for that is that we used assumptions which are better suited for divalent counterions. We assume constant counterion distribution function which is more valid in the case of divalent DNA/polyion than in the case of monovalent for which it changes more. Also the change is biggest for $x<0.2$ where potential $V$ changes most which causes that product of this functions $f(\mu)=\int_1^{1-x}Vq(x)dx$ could overestimate $\mu$. 
	
	Long history of studying counterion polarization brings many idea like existence of longitudinal and transverse polarizability\cite{fixman}, different electric potential of polymer ends as compared to the mid section\cite{odijk}. Both this ideas seem incorrect for long polymers $>>\ l/r_a$ in semidilute solutions. For a long time it is known that $L$ of semidilute solutions does not depend on polymer contour length $l$ while theory of longitudinal and transverse polarizability predict that it should. Ito et al\cite{ito} showed that $L$ of N=87 $(l/r_a=40)$ and N=3800 $(l/r_a=2300)$ Na-PSS are identical in semidilute solution while theory\cite{fixman} predict that total polarizabilities (absolute sum of longitudinal and transvers polarizabilities for both counterions and coions) for $(l/r_a=40)$ are smaller by 25\% as compared to $(l/r_a=180)$. And that is much smaller range as compared to range for which experiment did not show any change in $L$. Also based on mine and Ito's experimental work, in which $L$ does not change when polyion length is changed for more than 50 times, I suspect that effect of different potential on the ends of the polyion\cite{odijk} is not observable in semidilute solutions. So it is questionable if such effect should happen in semidilute solutions? I suspect this effect could be observable in dilute solutions in which distances between polyions $\xi$ are much larger than in semidilute solutions. Even more, theory predict that such effect could be expected only within $\kappa^{-1}$ distance from the edges of the polyion. This distance is 10-1500 and 10 times respectively smaller than the contour length of polydisperse DNA and PSS published in this works. So one should really not expect to observe such effects in experiments like this. Simulations also supports view that in semidilute solution one should not observe different electric potential. They found that anisotropy of the electric polarizability in salt-free solutions decreases with increasing polyion concentration and that this decrease is proportional to second or higher power of molecular weight\cite{hitoshi 1,hitoshi 2}. Another phenomena expected for infinitely dilute polyelectrolyte in excess of simple salt is polarizability of condensed counterions parallel to polyion\cite{manning}. All this complicated phenomena, longitudinal and transverse polarization, different electric potential of edges as compared to mid region can be avoid by carefully designing experiments, in particularly in experiment with long polyions in semidilute solutions. Maybe that is the reason why nobody before did not discover such universal behavior of polyions and that is also the reason why we manage to explain our data with a very simple theory/model. Maybe more theoretical effort should be put in order to first fully understand such relatively simple regime of long semidilute polyelectrolyte and then one should try to better understand more complicated regimes like dilute and short polyelectrolytes.	 
	
	The importance of this new interpretation is that one can verify polyelectrolyte potential. Knowing polyelectrolyte potential one can model polyelectrolyte properties. First on that list should be the osmotic pressure. Conductivity has probably more complex dependence on the potential but it should also be governed by potential.
	
	We took the most simple potential but there are also other suggested polyelectrolyte potentials\cite{hansen,adriana}. Only exact mathematical formulation of the average dipole moment would discriminate the one which best agrees to experimental data. Now the question remains about interpretation of $L$ in solutions with added salt on which both theoretical and experimental work is needed. Good theoretical foundation already exist\cite{naji,ray98,ray99}.

	\section{CONCLUSION}
	
	If one would know the polyelectrolyte potential many of its mysteries would be understood, so there is a great need for experimental method which can test the polyelectrolyte potential. Such method is dielectric spectroscopy. From 1990 researchers believe that characteristic length of counterions relaxing in $10^5Hz-10^7Hz$ covers distance between two polyions, i.e. distance that has the same value as correlation length $\xi$. If this is correct it should not depend on counterion valency but we showed that it does. Moreover $\xi$ measured by SAXS is 6 times longer than $L$ both of which suggest that $L$ does not measure $\xi$. So there is a need for new interpretation of $L$. With the simplest possible model of a polyion we have shown that DS measures average dipole moment of polyelectrolytes which depends only on one parameter; p-c potential. This means that indirectly DS probes the polyion potential. Our model better predicts measured data than old model. We would like to add that we perturb counterions by energy which is more than 1000 times smaller than $kT$ which means that we measure potential of "unperturbed state".
	
	\section{Associated Content}
	In supporting information we provide proofs that fitting parameters of higher frequency relaxation are precisely determined despite the electrode polarization which prevents precise determination of parameters of lower frequency relaxation. We also proofs that parameters of higher frequency relaxation are in accord with the Ito et. al data.\cite{ito}. At the end we argue that deviation of L for dilute samples as compared to semidilute is expected. This information is available free of charge via the Internet at http://pubs.acs.org/.

	\section{ACKNOWLEDGEMENTS}
	I wish to express my sincere thanks to my colleague Damir Vurnek. Without his help this article would be a pale shadow of a work that now is.

	\newpage


\begin{thebibliography}{99}
		
		
		
		
		\bibitem{naji} Naji,A.; Kandu\v{c}, M.; Forsmann, J.; Podgornik, R. Perspective: Coulomb fluids--weak coupling, strong coupling, in between and beyond. {\it The journal of Chemical Physics} {\bf 2013}, {\it 139}, 1509011-15090112.
		
		\bibitem{Schmitz93} Schmitz, K.~S. In {\it  Macroions in Solution and Colloidal Suspension}; VCH: New York, 1993
		
		\bibitem{oosawaknjiga71} Oosawa, F. In {\it Polyelectrolytes}, 1st ed.; Marcel Dekker: New York, 1971
		
		\bibitem{katchalsky12} Katchalsky, A. Problems in the Physics of Polyelectrolytes {\it Journal of Polymer Science} {\bf 1954}, {\it 12}, 159-184.
		
		\bibitem{katchalsky13} Lifson, S.; Katchalsky, A. The Electrostatic Free Energy of Polyelectrolyte Solution. II. Fully Stretched Macromolecules. {\it Journal of Polymer Science} {\bf 1954}, {\it 13}, 43-55. 
		
		\bibitem{ray98} Manning, G.S.; Ray, J. Counterion Condensation Revisited. {\it J. Biomol. Struct. Dynam.} {\bf 1998}, {\it 16}, 461-476.
		
		\bibitem{ray99} Ray, J.; Manning, G.S. Counterion and coion distribution functions in the counterion condensation theory of polyelectrolytes. {\it Macromolecules} {\bf 1999}, {\it 32}, 4588-4595.
		
		\bibitem{dobrynin} Dobrnin, A. V.; Rubinstein, M.; Theory of polyelectrolytes in solutions and at surfaces Andrey. {\it Prog. Polym. Sci.} {\bf 2005}, {\it 30}, 1049-1118.	
		
		\bibitem{deshkovski} Deshkovski, A.; Obukhov, S.; Rubinstein, M. Counterion Phase Transitions in Dilute Polyelectrolyte Solutions Alexander. {\it Phys. Rev. Lett.} {\bf 2001}, {\it 86}, 2341-2344.
		
		\bibitem{antypov} Antypov, D.; Holm, C. Optimal Cell Approach to Osmotic Properties of Finite Stiff-Chain Polyelectrolytes. {\it Phys. Rev. Lett.} {\bf 2006}, {\it 96}, 0883021-0883024.
		
		\bibitem{hansen} Hansen, P. L.; Podgornik, R.; Parsegian, V. A. Osmotic properties of DNA: Critical evaluation of counterion condensation theory. {\it Phys. Rev. E} {\bf 2001}, {\it 64}, 02190711-02190714.	
		
		\bibitem{kern} Kern, W. Über heteropolare Molekülkolloide. I. Die Polyacrylsäure, ein Modell des Eiweies 185. Mitteilung über hochpolymere Verbindungen. {\it Z. Physik. Chemie.} {\bf 1938}, {\it 181A}, 249-282. 
		
		\bibitem {kern2} Kern, W. Der osmotische Druck wässeriger Lösungen polyvalenter Säuren und ihrer Salze 215. Mitteilung über makromolekulare Verbindungen. {\it Z. Physik. Chemie.} {\bf 1939}, {\it A184}, 197-210.
		
		\bibitem{nagasawa} Nagasawa, M.; Kagawa, I. Colligative properties of polyelectrolyte solutions. IV. Activity coefficient of sodium ion. {\it J. Polymer Sci.} {\bf 1957}, {\it 25}, 61-76.
		
		\bibitem{alexandrowicz} Alexandrowicz, Z. The Correlation between Activities of Polyelectrolytes, Measured by the Light-Scattering and Osmotic Methods. {\it J. Polymer Sci.} {\bf 1959}, {\it 40}, 91-106.
		
		\bibitem{alexandrowicz2} Alexandrowicz Z. Results of Osmotic and of Donnan Equilibria Measurements in Polymethacrylic Acid-Sodium Bromide Solutions. Part II.  {\it J. Polymer Sci.} {\bf 1960}, {\it 43}, 337-349.
		
		\bibitem{manning69} Manning, G. S. Limiting Laws and Counterion Condensation in Polyelectrolyte Solutions I. Colligative Properties. {\it The Journal of Chemical Physics} {\bf 1969}, {\it 51}, 924-933.
		
		\bibitem{liao} Liao, Q.; Dobrynin, A. V.; Rubinstein, M. Molecular Dynamics Simulations of Polyelectrolyte Solutions : Osmotic Coefficient and Counterion Condensation. {\it Macromolecules} {\bf 2003}, {\it 36}, 3399-3410.
		
		\bibitem{ito} Ito, K.; Yagi, A.; Ookubo, N.; Hayakawa, R. Crossover behavior in high-frequency dielectric relaxation of linear polyions in dilute and semidilute solutions. {\it Macromolecules}, {\bf 1990}, {\it 23}, 857-862.
		
		\bibitem{deGennes76} de Gennes, P.~G.; Pincus, P.; Velasco R.~M.; Brochard, F. Remarks on polyelectrolyte conformation. {\it Journal de Physique (Paris)} {\bf 1976}, {\it 37}, 1461-1473.
		
		\bibitem{grgicin2013} Grgi\v{c}in, D.; Dolanski Babi\'{c}, S.; Ivek, T.; Tomi\'{c}, S.; Podgornik, R. Effect of magnesium ions on dielectric relaxation in semidilute DNA aqueous solutions. {\it Phys. Rev. E} {\bf 2013}, {\it 88}, 052703
		
		\bibitem{salamon} Salamon, K.; Aumiler, D.; Pabst, G.; Vuleti\'{c}, T. Probing the Mesh Formed by the Semirigid Polyelectrolytes. {\it Macromolecules} {\bf 2013}, {\it 46}, 1107-1118.
		
		\bibitem{epl08} Tomi\'{c}, S.; Dolanski Babi\'{c}, S.; Ivek, T.; Vuleti\'{c},  T.; Kr\v{c}a, S.; Livolant, F.; Podgornik, R. Short-fragment Na-DNA dilute aqueous solutions: fundamental length scales and screening. {\it EPL} {\bf 2008,} {\it 81}, 680031-680035.
		
		\bibitem{record} Record, M. T. Effects of Na+ and Mg++ ions on the helix-coil transition of DNA. {\it Biopolymers} {\bf 1975}, {\it 14}, 2137-2158.
		
		\bibitem{combet} Combet, J.; Rawiso, M.; Rochas, C.; Hoffmann, S.; Boue, F. Structure of polyelectrolytes with mixed monovalent and divalent counterions: SAXS measurements and Poisson-Boltzmann analysis. {\it Macromolecules} {\bf 2011}, {\it 44}, 3039-3052.
		
		\bibitem{pre07} Tomi\'{c}, S.; Dolanski Babi\'{c}, S.; Kr\v{c}a, S.; Ivankovi\'{c}, D.; Gripari\'{c}, L.; Podgornik, R. Dielectric relaxation of DNA aqueous solutions. {\it Physical Review E} {\bf 2007}, {\it 75}, 0219051-02190513.
		
		\bibitem{norio} Ookubo, N.; Hirai, Y.; Ito, K.; Hayakawa, R. Anisotropic counterion polarizations and their dynamics in aqueous polyelectrolytes as studied by frequency-domain electric birefringence relaxation spectroscopy. {\it Macromolecules} {\bf 1989}, {\it 22}, 1359-1366.
		
		\bibitem{saif} Saif, B.; Mohr, R. K.; Montrose, C. J.; Litovitz, T. A. On the mechanism of dielectric relaxation in aqueous DNA solutions. {\it Biopolymers} {\bf 1991}, {\it 31}, 1171-1180.
		
		\bibitem{litovitz} Penafiel, L. M.; Litovitz, T. High frequency dielectric dispersion of polyelectrolyte solutions and its relation to counterion condensation. {\it The Journal of Chemical Physics} {\bf 1992}, {\it 97}, 559-567. 
		
		\bibitem{lee} Lee, R. S.; Bone, S. Dielectric studies of chain melting and denaturation in native DNA. {\it Biochimica et Biophysics Acta} {\bf 1998}, {\it 1397}, 316-324.
		
		\bibitem{gabriel} Gabriel, C.; Grant, E. Dielectric behavior of frozen DNA in solution. {\it Bioelectromagnetics} {\bf 1999}, {\it 20}, 40-45.
		
		\bibitem{hayakawa} Nagamine, Y.; Ito, K.; Hayakawa R. Low- and high-frequency relaxations in linear polyelectrolyte solutions with different counter-ion species. {\it Langmuir} {\bf 1999}, {\it 15}, 4135-4138.
		
		\bibitem{bakewell} Bakewell, D. J.; Ermolina, I.; Morgan, H.; Milner, J.; Feldman, Y. Dielectric relaxation measurements of 12 kbp plasmid DNA. {\it Biochimica et Biophysica Acta} {\bf 2000}, {\it 1493}, 151-158. 
		
		\bibitem{omori} Omori, S.; Katsumoto, Y.; Yasuda, A.; Asami, K. Dielectric dispersion for short double-strand DNA. {\it Physical Review E} {\bf 2006}, {\it 73}, 0509011-0509014.
		
		\bibitem{katsumoto} Katsumoto, Y.; Omori, S.; Yamamoto, D.; Yasudu, A. Dielectric dispersion of short single-stranded DNA in aqueous solutions with and without added salt. {\it Physical Review E} {\bf 2007}, {\it 75}, 011911-011919.
		
		\bibitem{elena} Ermilova, E.; Bier, F. F.; H{\"o}lzel, R. Dielectric measurements of aqueous DNA solutions up to 110 GHz. {\it Phys. Chem. Chem. Phys.} {\bf 2014}, {\it 16}, 11256-11264.
		
		\bibitem{Grgicin2018}  Grgi\v{c}in, D.; Vurnek. In preparation
		
		\bibitem{bloomfield} Bloomfield, V.A.; Crothers, D. M.; Jr. Tinocco, I. {\it Nucleic Acids}, University Science Books, Sausalito (2000)
		
		\bibitem{physicab} Tomi\'{c}, S.; Grgi\v{c}in, D.; Ivek, T.; Vuleti\'{c}, T.; Dolanski Babi\'{c}, S.; Podgornik R. Dynamics and structure of biopolyelectrolytes in repulsion regime characterized by dielectric spectroscopy. {\it Physica B} {\bf 2012}, {\it 407}, 1958-1963.
		
		\bibitem{vuletic10} Vuleti\'{c}, T.; Dolanski Babi\'{c}, S.; Ivek, T.; Grgi\v{c}in, D.; Tomi\'{c}, S.; Podgornik, R. Structure and dynamics of hyaluronic acid semidilute solutions: A dielectric spectroscopy study. {\it Phys. Rev. E} {\bf 2010}, {\it 82}, 0119221-01192210.
		
		\bibitem{Lide} Vanysek, P. In {\it CRC Handbook of Chemistry and Physics}, 88 ed.; Lide, D. R. Ed.; CRC Press: Boca Raton, 2007; p. 5-76. 	
		
		\bibitem{hazu} Tomi\'{c}, S.; Grgi\v{c}in,  D.; Vuleti\'{c}, T.; Dolanski Babi\'{c}, S.; Ivek, T.; Podgornik, R. DNA in aqueous solutions with repulsive interactions: structure determined on the basis of dielectric spectroscopy measurements, {\it Bioinformatics and biological physics: proceedings of the scientific meeting, edited by Vladimir Paar, Zagreb, Croatian Academy of Sciences and Arts, Committee for Bioinformatics and Biological Physics of Department for Mathematical, Physical and Chemical Sciences} {\bf 2013}, 159-177.
		
		\bibitem{Bordi}	Bordi, F.; Cametti, C.; Colby, R. H. Dielectric spectroscopy and conductivity of polyelectrolyte solutions
		{\it J. Phys.: Condens. Matter} {\bf 2004}, {\it 16}, R1423.
		
		\bibitem{Schwan68} H.\ P.\ Schwan, and C.\ D.\ Ferris, Four Electrode Null Techniques for Impedance Measurement with High Resolution {\it Rev.\ Sci.\ Instrum.}\ {\bf 1968}, {\it 39}, 481
		
		\bibitem{Holling} Hollingsworth, A. D.; Saville, D. A. A broad frequency range dielectric spectrometer for colloidal suspensions: cell design, calibration, and validation {\it J. Colloid Interface Sci.} {\bf 2003}, {\it 257}, 65.
		
		\bibitem{Roldan} Roldan-Toro, R.; Sollier, J. D. Wide-frequency-range dielectric response of polystyrene latex dispersions {\it J.\ Colloid Interface Sci.} {\bf 2004}, {\it 274}, 76.
		
		\bibitem{Klein} Klein, R. J.; Zhang, S. Dou, S.; Jones B. H.; COlby, R. H. and R, J. Modeling electrode polarization in dielectric spectroscopy: Ion mobility and mobile ion concentration of single-ion polymer electrolytes {\it The Journal of Chemical Physics} {\bf 2006}, {\it 144903}, 1449031-144907
		
		\bibitem{Davey} Davey,	C. L.; Markx, G. H.; Kell,D. B. Substitution and spreadsheet methods for analyzing dielectric spectra of biological systems {\it Eur. Biophys. J.} {\bf 1990} {\it 18}, 255.
		
		\bibitem{Grosse} Grosse, C.; Tirado, M.; Pieper, W.; Pottel, R. Broad Frequency Range Study of the Dielectric Properties of Suspensions of Colloidal Polystyrene Particles in Aqueous Electrolyte Solutions {\it J. Coll. Interface Sci.} {\bf 1998}, {\it 205}, (1998).
		
		\bibitem{ito komora} Hayakawa, R.; Kanda, H.; Sakamoto, M.; Wada, Yasaku. New Apparatus for Measuring the Complex Dielectric Constant of a Highly Conductive Material {\it Japanese Journal of Applied Physics} {\bf 1975}, {\it 14}, 2039-2052
		
		\bibitem{ida} Dela\'{c} Marion, I.; Grgi\v{c}in, D.; Salamon, K.; Bernstorff, S.; Vuleti\'{c}, T. Polyelectrolyte Composite: Hyaluronic Acid Mixture with DNA. {\it Macromolecules} {\bf 2015}, {\it 48}, 2686-2696.
		
		\bibitem{deserno} Deserno, M.; Holm, C.; May, S. Fraction of Condensed Counterions around a Charged Rod: Comparison of Poisson-Boltzmann Theory and Computer Simulations. {\it Macromolecules} {\bf 2000}, {\it 33}, 199-206. 
		
		\bibitem{adriana} Nicasio-Collaza, L. A.; Delgado-Gonzalez, A.; Hernandez-Lemus, E.; Castaneda-Priego, R. Counterion accumulation effects on a suspension of DNA molecules: Equation of
		state and pressure-driven denaturation. {\it The Journal of Chemical Physics} {\bf 2017}, {\it 146}, 1649021-1649029.
		
		\bibitem{fixman} Fixman, M.; Jagannathan, S. Electric and convective polarization of the cylindrical macroions. {\it The Journal of Chemical Physics}, {\bf 1981}, {\it 75}, 4048-4059.
		
		\bibitem{odijk} Odijek, T; Impact of nonuniform counterion condensation on the growth of linear charged micellec. {\it Physica A} {\bf 1991}, {\it 176}, 201-205.
		
		\bibitem{hitoshi 1} Washizu, H; Kikuchi, K.  Electrical Polarizability of Polyelectrolytes in Salt-free Aqueous Solution. {\it J. Phys. Chem. B} {\bf 2002}, {\it 106}, 11329-11342.
		
		\bibitem{hitoshi 2} Washizu, H; Kikuchi, K.  Electrical Polarizability of DNA in Aqueous Salt Solution {\it J. Phys. Chem. B} {\bf 2006}, {\it 110}, 2855-2861.
		
		\bibitem{manning} Manning, G. S. Limiting Laws and counterion condensation in polyelectrolyte solutions V. Further development of the chemical model. {\it Biophysical Chemistry} {\bf 1978}, {\it 9}, 65-70.
		
	\end{thebibliography}
\end{document}